\documentclass[letterpaper]{pac97}    
\usepackage{epsfig}    
\begin{document}

\title{THE ALGEBRAIC  RECONSTRUCTION  TECHNIQUE 
(ART)\thanks{Work performed under the auspices of 
the U. S. Department of Energy. }}
\author{D.Raparia,  J.Alessi,  and A.Kponou\\
 AGS Department, Brookhaven  National Lab,  Upton, NY 11973, USA  }

\maketitle

\begin{abstract}  
Projections of charged particle beam current density (profiles) 
are frequently used 
as a measure of beam position and size. In conventional practice only
two projections, usually horizontal and vertical, are measured. This puts 
a severe limit on the detail of information that can be 
achieved. A third projection provides a significant improvement. 
The Algebraic Reconstruction Technique (ART)
 uses three or more
projections to reconstruct 3-dimensional density profiles.  
At the 200 MeV H- linac, we have used this
technique to measure  beam density, and it has proved 
very helpful, especially in helping determine 
if there is any coupling present in x-y phase space.
We will present examples of measurements of 
current densities using this technique.

\end{abstract}

\section{Introduction}
In Computed Tomography (CT), three dimensional reconstruction techniques from
projection have been used for many years in radiology. The two dimensional
Fourier transform is the most commonly used algorithm in radiology. In
this technique a large number of projections at uniformly distributed angles
around the subject are required for reconstruction of the image. In the field
of accelerator physics, one expects that the relatively simple charged particle
beam distributions can be reconstructed from a small number of projections. 
In conventional practice only
two projections, usually horizontal and vertical, are measured. This puts 
a severe limit on the level of detail that can be 
achieved. The Algebraic Reconstruction Technique (ART) 
introduced by Gordan, Bender and Herman \cite{gordan} 
 uses three or more projections to reconstruct the 
2-dimensional beam density distribution.  They have shown that the
improvement in the quality of the reconstruction is pronounced when a 
third projection is added, but additional projections add 
much less to the reconstruction quality.

\section{Algebraic Reconstruction Technique (ART)}

 The ART algorithms have a simple intuitive basis.
Each projected density is thrown back across the reconstruction space in which
the densities are iteratively modified to bring each reconstructed projection 
into agreement with the measured projection. Assuming that the pattern being
 reconstructed is enclosed in a square space of  n x n 
array of small pixels, $\rho_{j}\left( j=1,\ldots ,n^2\right)$ is grayness
or density number, which is uniform within the pixel but different from
 other pixels.
A ``ray'' is a region of the square space which lies between two parallel lines.
The weighted ray sum is the total grayness of the reconstruction 
figure within the ray. The projection at a given angle is 
then the sum of non-overlapping,
equally wide rays covering the figure. The ART algorithm consists of altering
the grayness of each pixel intersected by the ray in such a way as to make the 
ray sum agree with the corresponding element of the measured projection. 
Assume \bf{P} \rm is a matrix of m x n$^{2}$ and the m component column vector
\bf{R}. \rm Let $p_{i,j}$ denote the (i,j)th 
element of \bf{P} \rm, and $R_i$ denote the ith ray of the 
reconstructed projection vector \bf{R}\rm.
 For $1 \leq i \leq m$, N$_i$ is number of pixels under projection ray R$_i$,
defined as
 $ N_{i} = \sum_{j=1}^{n^2} p_{i,j}^{2}$.
ART is an iterative method.
 The density number $ \rho_j^q $ denotes the value of $ \rho_j $ after q
iterations.
 After q iterations the intensity of the ith reconstructed projection ray is 
 $$ R_i^q = \sum_{j=1}^{n^2} p_{i,j} \rho_j^q , $$
and the density in each pixel is 
  $$\rho^{\sim q+1}_j= \rho_j^q + p_{i,j} {R_i - R_i^q \over N_i}
 ~~~~\mbox{with starting value}~~\rho^{\sim 0}_j=0 $$
where R$_i$ is the measured projection ray and,
\[ i= \left \{ \begin{array}{l}
              \mbox{m, if (q+1) is divisible m} \\
	      \mbox{the remainder of dividing (q+1)by m, otherwise}
		\end{array}
		\right. \] 
and,
    \[  \rho_j^q  = \left \{ \begin{array}{lll}
                               0,                   & \mbox{if $ \rho^{\sim q } \leq 0 $} \\
                                \rho^{\sim q}_j , & \mbox{if $0 \leq \rho^{ \sim q}_j \leq 1$} \\
			       1,                   & \mbox{if $ \rho^{\sim q}_j \geq 1$}
                               \end{array} 
                             \right. \]
This algorithm is known as fully constrained ART.

It is necessary to determine when an iterative algorithm has converged
to a solution which is optimal according to some criterion. Various criteria
for convergence have been devised.
 The discrepancy between the measured and calculated
projection elements is
$$D^q \equiv \left\{ \frac{1}{m} \sum_{i=1}^m  \frac{\left( R_i - R_i^q \right)^2}{N_i} \right\}^{\frac{1}{2}} ,$$
and the nonuniformity or variance of constructed figure is
$$ V^q \equiv \sum_j \left( \rho_j^q - \overline{\rho} \right)^2 ,$$
and  the entropy constructed figure is 
$$ E^q \equiv \frac{-1}{2 \log n} \sum_j \left( \frac{\rho_j^q}{\overline{\rho}} \right)
 \log \left( \frac{\rho_j^q}{\overline{\rho}} \right) .$$
$D^q$ tends to zero, $V^q$ to a  minimum and $S^q$ to 
a maximum with increasing q. For a known  
test pattern ($\rho^t_{i,j}$), the Euclidean Distance is define as
$$ s^q \equiv \sqrt{\frac{1}{n^2} \sum_j \left( \rho^q_j - \rho^t_j \right)^2}. $$

\section{ Test Figure}

It is instructive to test the reconstruction capabilities of ART
with two to four views by using projections from a known test figure.
In the following example, we have used an x-y coupled (about
18$^{\circ}$ )two-dimensional gaussian enclosed in a square space of 100 x 100
array with $\sigma_x=5$ and $\sigma_y=20$. We have used a ray width in the
45$^{\circ}$ and 135$^{\circ}$ projection as $\sqrt{2}$ times 
of ray width in x or y projection, making number of ray in 
each projection same namely 100. 
Fig. 1 shows the test figure and reconstructed test figure from
two projections. Fig. 2 shows reconstructed test figure from
three and four projections.

\begin{figure}[htb]
\hspace{-5mm}
\scalebox{0.6}{
\rotatebox{90}{\mbox{\epsfig{file=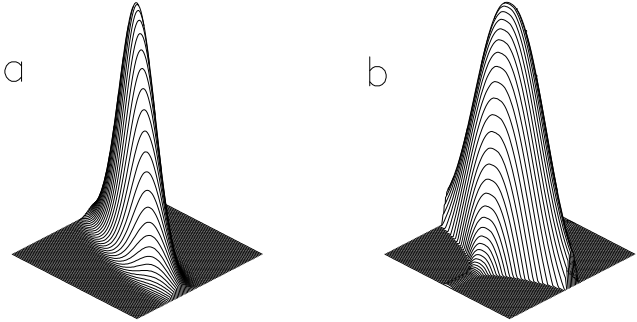,width=0.85\columnwidth}}}}
\vspace{-1.5mm}
\caption{ (a) Original test figure and (b) reconstructed test figure from
two projections.}
\vspace{3.mm}
\end{figure}

\begin{figure}[htb]
\hspace{-5mm}
\scalebox{0.6}{
\rotatebox{90}{\mbox{\epsfig{file=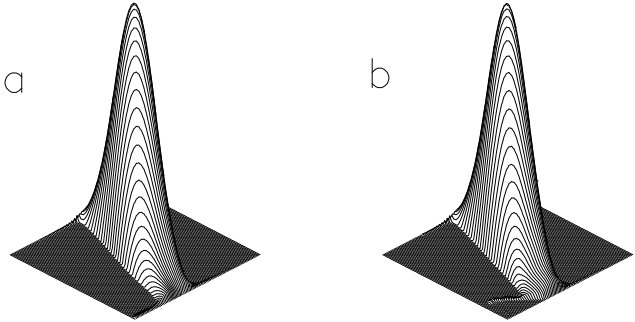,width=0.85\columnwidth}}}}
\vspace{1.5mm}
\caption{ Reconstructed test figures from (a) three and (b) four projections.}
\vspace{3.mm}
\end{figure}
Fig. 3 shows the contours of Figures 1 and 2.
\begin{figure}[htb]
\hspace{-0mm}
\scalebox{0.92}{
\rotatebox{90}{\mbox{\epsfig{file=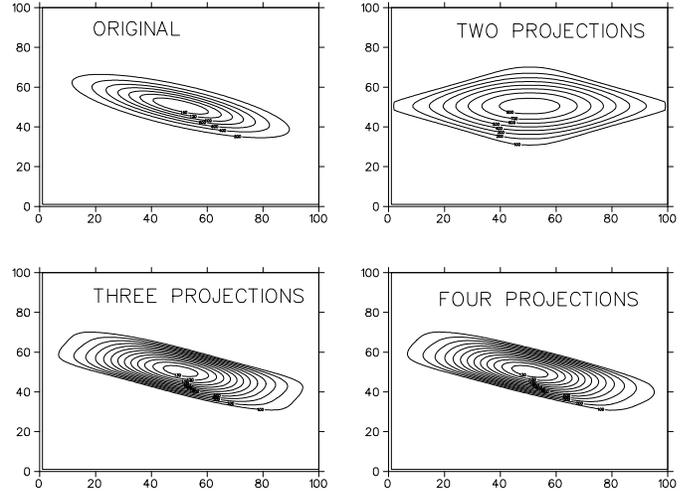,width=0.85\columnwidth}}}}
\vspace{+1.5mm}
\caption{Contour plots of test figure and reconstructed figures 
with two, three and four projections.}
\vspace{+3.mm}
\end{figure}
\begin{table}[htb]
 \begin{center}
	\caption{ The convergence criteria discrepancy (D), 
	variance (V),  the entropy (E) and the 
	Euclidean Distance (s) for two, three and four projections.}
\vspace{+5mm}
	  \begin{tabular}{|l|l|l|l|l|l|l|}
	\hline
                						&
	        2 Proj.					&
		3 Proj.				&
  		4 Proj.				\\
	  \hline Iteration No					&
		67						&
		1426						&
		1083						\\
	\hline  Time (sec)					&
		201						&
		6393						&
		6534						\\
	\hline Discrepancy					&
		1.0 10$^{-6}$					&
		1.0 10$^{-6}$					&
		1.0 10$^{-6}$					\\
	\hline Variance					&
		4.8 10$^{-8}$					&
		1.3 10$^{-8}$					&
		1.3 10$^{-8}$					\\
	\hline Entropy						&
		1.9 10$^{+3}$					&
		2.3 10$^{+3}$					&
		2.4 10$^{+3}$					\\
	\hline E. Distance 				&
		1.5 10$^{-4}$					&
		4.6 10$^{-5}$					&
		4.6 10$^{-5}$					\\
	\hline
  \end{tabular}
 \end{center}
\end{table}
\begin{figure}[htb]
\hspace{-0mm}
\scalebox{0.85}{
\rotatebox{90}{\mbox{\epsfig{file=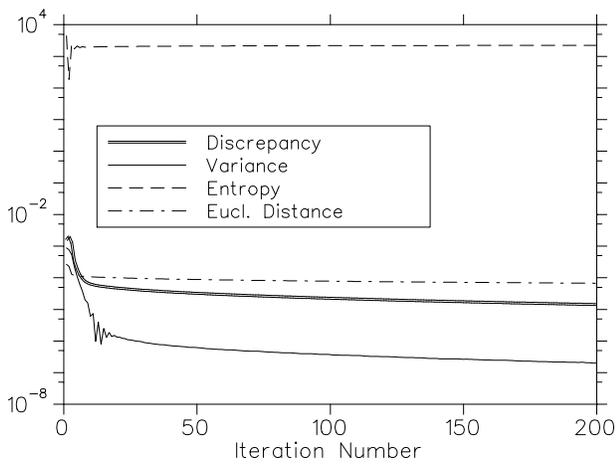,width=0.85\columnwidth}}}}
\vspace{1.5mm}
\caption{The discrepancy (D), variance (V),  
 entropy (E) and the Euclidean Distance (s) 
as a function of iteration number for case of 
three projections. The convergence criteria  was 
if discrepancy is less the 10$^{-6}$.}
\vspace{-3.mm}
\end{figure}
It is clear from  Fig. 3 that  two projections are not enough for catching 
the coupling. The accuracy of the reconstructed figure from
 four projection is slightly better than three projections. 
Fig. 4 shows the discrepancy (D), variance (V),  
the entropy (E) and the Euclidean Distance (s) 
as a function of iteration number for case of 
three projections. The convergence 
criteria  was if discrepancy is less than 10$^{-6}$.
Table 1 show the numerical values of discrepancy (D), 
variance (V),  the entropy (E) and the Euclidean Distance (s) 
for two, three and four projections.

\section{ Beam Density Measurement}
There are stepping wire profile scanners at 13 
locations throughout the 200 MeV linac and transport lines.  These 
scanners are mounted at a 45$^{\circ}$ angle with 
respect to horizontal, and single horizontal and 
vertical wires are stepped through the bea
We have added a third wire at 45$^{\circ}$ to horizontal in 
two of the scanners, one in the 750 keV line \cite{linac96a} and 
one in the 200 MeV BLIP \cite{linac96b} transport line.
Fig. 5 shows a schematic of the scanner with three wires.
\begin{figure}[htb]
\vspace{5mm}
\hspace{-0mm}
\scalebox{1.0}{
\rotatebox{0}{\mbox{\epsfig{file=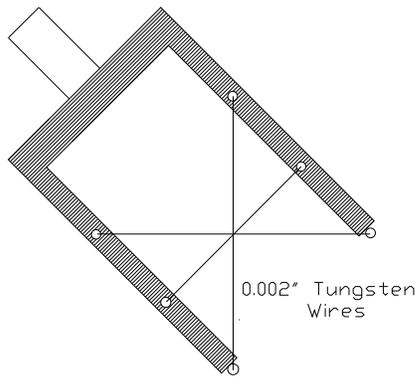,width=0.85\columnwidth}}}}
\vspace{-1.5mm}
\caption{Schematic of the scanner with three wires.}
\vspace{-3.mm}
\end{figure}
Fig. 6   shows the reconstructed density distributions at 750 keV line.
There is no x-y coupling in the 750 keV line.
\begin{figure}[htb]
\hspace{-0mm}
\scalebox{0.70}{
\rotatebox{90}{\mbox{\epsfig{file=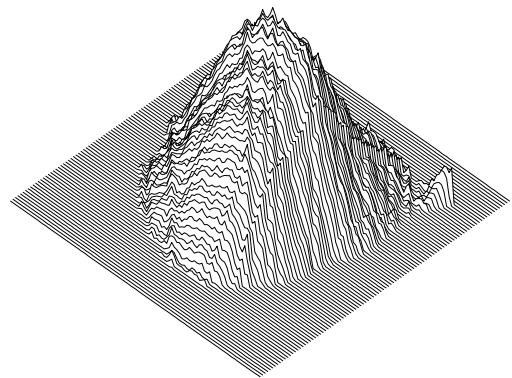,width=0.85\columnwidth}}}}
\vspace{-1.5mm}
\caption{Reconstructed 3D density distribution in the 750 keV line using ART.}
\vspace{-3.mm}
\end{figure}
Fig. 7 shows beam density contour plots in the  BLIP line.  
The x-y coupling is clearly seen. This coupling could come from one  or
more rotated quadrupoles or vertical beam offset in a dipole. In the presence
of x-y coupling,  the usual technique of emittance measurement
from profiles at   three or
more  locations will not work.
\begin{figure}[htb]
\hspace{-0mm}
\scalebox{0.70}{
\rotatebox{90}{\mbox{\epsfig{file=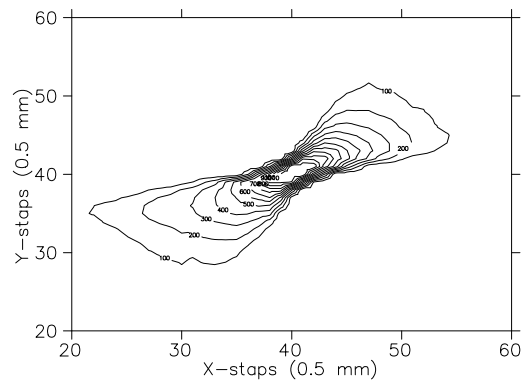,width=0.85\columnwidth}}}}
\vspace{-1.5mm}
\caption{Reconstructed contour plot using ART in the  BLIP line, 
showing x-y coupling.}
\vspace{-3.mm}
\end{figure}
Figure 8 compares the measured and reconstructed projections in the BLIP line.
\begin{figure}[htb]
\hspace{-0mm}
\scalebox{0.70}{
\rotatebox{90}{\mbox{\epsfig{file=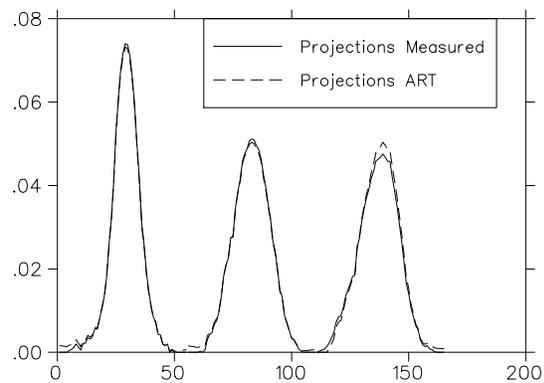,width=0.85\columnwidth}}}}
\vspace{-1.5mm}
\caption{Beam projection measured and reconstructed 
on X, Y, and 45$^{\circ}$  planes at BLIP line.}
\vspace{-3.mm}
\end{figure}

\end{document}